\documentclass[]{article}
\usepackage{graphicx}
\def\title#1{\begin{centering}\Large\bf #1 \\[5mm]\end{centering}}
\def\author#1{\centerline{#1}}
\def\address#1{\begin{centering}\small #1 \\[5mm]\end{centering}}

\begin {document}
\title{Atom Interferometry in a Vertical Optical Lattice }
\author{G. Modugno, E. de Mirandes, F. Ferlaino, H. Ott, G. Roati, M. Inguscio}
\address{LENS and Dipartimento di Fisica, Universit\`a di Firenze,
and INFM, Via Nello Carrara 1, 50019 Sesto Fiorentino, Italy}

\begin{abstract} We have studied the interference of degenerate
quantum gases in a vertical optical lattice. The coherence of the
atoms leads to an interference pattern when the atoms are released
from the lattice. This has been shown for a Bose-Einstein
condensate in early experiments. Here we demonstrate that also for
fermions an interference pattern can be observed provided that the
momentum distribution smaller then the recoil momentum of the
lattice. Special attention is given to the role of interactions
which wash out the interference pattern for a condensate but do
not affect a spin polarized Fermi gas, where collisions at ultra
cold temperatures are forbidden. Comparing the interference of the
two quantum gases we find a clear superiority of fermions for
trapped atom interferometry.
\end{abstract}

\section{Introduction}
Atom interferometry \cite{Berman1997a} lies at the heart of
quantum mechanics because it directly reveals the wave nature of
massive particles. It is also important with respect to technical
applications because it can be used for precision measurements of
forces and rotations \cite{Peters1999a,McGuirk2002a,Gupta2002a}.
To observe the interference, an atomic beam or a trapped atomic
sample is subjected to a periodic scatterer --- often realized
with an optical lattice. The lattice probes the coherence of the
atoms between neighboring lattice sites, thus leading to
constructive or destructive interference, depending on the
relative phase between the different paths. On a distant screen,
an interference pattern is observed if the width of the initial
momentum spread of the atoms in the direction of the lattice
$\delta q$ is smaller than the recoil momentum of the lattice
$\pi/d$, where $d$ is the lattice spacing. For a thermal beam this
condition can be achieved by using apertures which filter out a
coherent part of the beam. For a trapped sample however, the
temperature of the atoms must be reduced below the recoil
temperature. An optical lattice created by two counter propagating
laser beams has a lattice spacing of $d=\lambda/2$, where
$\lambda$ is the wavelength of the laser. The recoil temperature
which is associated with the recoil momentum is in general below 1
$\mu$K. Consequently, the experimental conditions for observing
interference with trapped atoms are similar to the requirements
for achieving Bose-Einstein condensation. It is for this reason,
that the first experimental observation of the interference of
trapped atoms was made with a Bose-Einstein condensate
\cite{Anderson1998a}. Being a completely coherent sample a
condensate shows high contrast and good visibility. However the
required momentum condition can also be achieved with other atomic
samples such as thermal clouds \cite{BenDahan} or degenerate Fermi
gases. Indeed our experimental results presented in this work
prove that the observation of atomic interference within a trapped
sample is not restricted to Bose-Einstein condensates. On the
contrary atomic Fermi gases exhibit an even superior performance
because the initially good visibility of the interference pattern
of a condensate is counterbalanced by a washing out of the fringes
for longer holding times.

The work is organized as follows: in the first part we introduce
the theoretical concepts to describe particles in a periodic
potential which is subjected to an external force. In the second
part we introduce our experimental setup, the loading procedure of
the atoms in the lattice and show how the interference pattern can
be detected. In the last part we present our results and discuss
possible extensions of our work for future applications in
precision interferometry.

\begin{figure}
\begin{center}
\includegraphics[width=8cm]{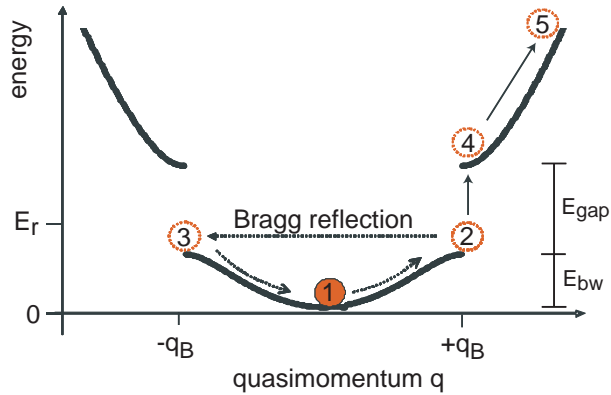} \caption{\label{fig1}
Dispersion relation: energy of the quasi momentum states of a
periodic potential, presented in an extended-zone scheme. The
bandgap $E_{\mathrm{gap}}$ and the band width $E_{\mathrm{bw}}$
are indicated on the right. A particle that is initially prepared
at rest (position 1, $q=0$) changes its quasi momentum under a
constant external force $F$: $\dot{q}=F/\hbar$. The quasi momentum
increases linearly until it reaches the Bragg momentum $q_{\mathrm
B}=\pi/d=2\pi/\lambda$ (position 2). For a sufficiently high
lattice it is Bragg reflected and reappears at the opposite side
of the Brillouin zone (position 3, $-q_{\mathrm B}$). It then
scans anew the first Brillouin zone, thus performing a periodic
motion in quasi momentum and in real space. In case of a shallow
lattice there is a finite probability to undergo a Zener
tunnelling transition from position 3 to position 4 corresponding
to a band transition from the first to the second band. The
particle is not reflected at the band edge and gains more quasi
momentum (position 5). Subsequent tunnelling into higher bands is
more and more likely and the particles motion can follow the
external force finally being released from the lattice.}
\end{center}
\end{figure}

\section{Theoretical Concepts}\label{section1}
Particles in a homogeneous periodic potential are described with
Bloch states. They extend over the whole lattice and are
characterized by the quasi momentum $q$ and the band index $n$.
The dispersion relation of the Bloch states for a sinusoidally
potential is shown in Fig.\,\ref{fig1} in an extended-zone-scheme:
it shows the typical band structure with the band gap
$E_{\mathrm{gap}}$ at the recoil energy
$E_{\mathrm{r}}=\hbar^2(2\pi/\lambda)^2/2m$, where $m$ is the mass
of the particle. For simplicity we only consider one dimension.
Without an external force, a particle which is prepared in a quasi
momentum state $q$ and a band $n$ will remain in this state and
propagate with a velocity which is given by the first derivative
of the dispersion relation
\begin{equation}\label{eq1}
v=\frac{1}{\hbar}\frac{\partial E(q)}{\partial q}.
\end{equation}
In the presence of a constant external force $F_{\mathrm{ext}}$,
the dynamics of the particle can be described within a
semiclassical approximation \cite{Ashcroft} and the quasi momentum
evolves according to
\begin{equation}\label{eq2}
\dot{q}=\frac{F_{\mathrm{ext}}}{\hbar}.
\end{equation}
When the quasi momentum of a particle which is prepared in the
lowest band equals the Bragg momentum
$q_{\mathrm{B}}=\pi/d=2\pi/\lambda$ there are two options: if the
band gap is large it stays confined within its band and is Bragg
reflected from the lattice (Fig.\,\ref{fig1}). It reappears on the
opposite site of the first Brillouin zone ($\pm q_{\mathrm{B}}$)
and scans anew the Brillouin zone. The temporal evolution of the
quasi momentum is hence a periodic sawtooth function. These
dynamics are known as Bloch oscillations \cite{BenDahan}. The
angular frequency $\omega_{\mathrm{BO}}$ and the amplitude $a$ are
given by
\begin{eqnarray}\label{BO}
\omega_{\mathrm{BO}}=\frac{F_{\mathrm{ext}}\lambda}{2\hbar}=\frac{mg\lambda}{2\hbar}\\
a=\frac{E_{\mathrm{bw}}}{2F_{\mathrm{ext}}}=\frac{E_{\mathrm{bw}}}{2mg},
\end{eqnarray}
where $E_{\mathrm{bw}}$ is the band width. The second identity
gives the explicit expression for the gravitational force
$F_{\mathrm{ext}}=mg$. The second option is a Zener tunnelling
process into the second band, whose probability depends strongly
on the chosen parameters of the potential and can be tuned by
changing the lattice depth.

\section{Experimental Setup}
In the experiment we employ a mixture of bosonic $^{87}$Rb and
fermionic $^{40}$K atoms. Both species are prepared in a
magneto-optical trap and subsequently transferred in a magnetic
trapping potential with trapping frequencies of
$\omega_a=2\pi\times16$ Hz and $\omega_r=2\pi\times200$ Hz in the
axial and radial direction (for $^{87}$Rb). During the forced
evaporation of $^{87}$Rb the fermions are sympathetically cooled
and we can bring both species simultaneously to quantum degeneracy
\cite{Roati2002a}. By removing all bosons at the end of the
evaporation ramp we can produce a pure Fermi gas of about
$3\times10^4$ atoms spin polarized in the $F=9/2,m_{\mathrm
F}=9/2$ state. The typical temperature is $T=0.3T_{\mathrm F}$,
where $T_{\mathrm F}=330$ nK is the Fermi temperature. A pure
Bose-Einstein condensate can be produced with about $10^5$ atoms
in the $F=2, m_{\mathrm F}=2$ state. We then switch on
adiabatically a lattice created by a retroreflected laser beam
aligned along the vertical direction.

\begin{figure}
\begin{center}
\includegraphics[width=10cm]{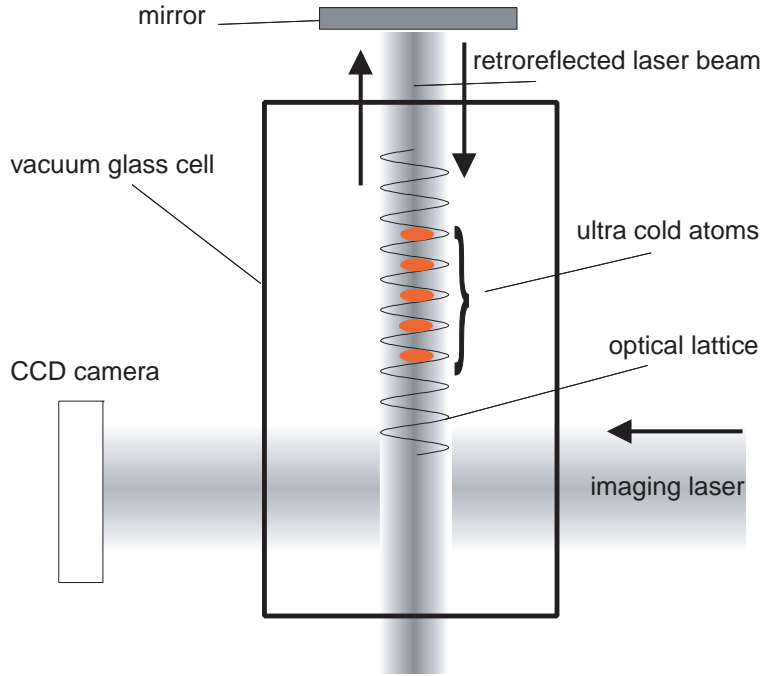} \caption{\label{fig2}
Sketch of the experimental setup. The atoms are loaded in a
vertical optical lattice which is created by a retroreflected,
off-resonant laser beam. After a given evolution time, the atoms
are released from the potential and fall freely in the
gravitational potential. After an expansion of 8-20 ms the atoms
are imaged with a resonant imaging laser.}
\end{center}
\end{figure}

The wavelength of the lattice laser is far detuned to the red of
the optical atomic transitions ($\lambda=873$ nm) to avoid photon
scattering. The depth of the potential can be adjusted in the
range of $U_\mathrm{K}=1-4E_{\mathrm r}$ for the fermions and
$U_\mathrm{Rb}=1-10E_{\mathrm r}$ for the bosons. Note, that the
recoil energy $E_\mathrm{r}=\hbar^2k^2/2m$ differs by factor of 2
for the two species due to the different mass. Because both, the
Fermi temperature and the critical temperature for Bose-Einstein
condensation are comparable to the recoil energy divided by the
Boltzmann constant, the atoms are loaded mostly in the first band
of the lattice potential. The Gaussian profile of the lattice
beams provides the confinement in the horizontal plane. Using
different intensities for the two beams we obtain a radial depth
of the optical potential of about 10 $E_\mathrm{r}$, with a
typical trap frequency of $2\pi\times30$ s$^{-1}$. Due to the
harmonic radial confinement every quasi momentum state has an
additional harmonic oscillator ladder for each of the two radial
dimensions. However, the Hamiltonian of the 3D system can be
separated in its three coordinates \cite{separation} and the
motion in the direction of the lattice is decoupled from the
motion in the radial direction. Therefore the three dimensional
nature of our setup does not change the equation of motion
(\ref{eq2}). A sketch of the experimental setup is shown in
Fig.\,\ref{fig2}.

To study the interference we suddenly switch off the magnetic trap
and let the atoms evolve in the combined lattice and gravitational
potential. In our range of lattice depths only the atoms in the
first band are trapped against gravity, while the small fraction
in excited bands falls down freely. After a given evolution time
we release the atoms from the trap and probe the cloud by
absorption imaging after 8 ms (20 ms) of free expansion for the
fermions (bosons). The interference of the atomic cloud can be
made visible in different ways. The first method, where the
potential height is lowered to a small value until the atoms drop
out of the potential via Zener Tunnelling processes leads to a
continuous output. For our setup the probability for such a
process is given by $P=e^{-\lambda E_{\mathrm{gap}}^2/8\hbar^2g}$,
where $g$ is the earth acceleration, and the band gap
$E_{\mathrm{gap}}$ can be derived from the lattice depth
$U=sE_{\mathrm{r}}$. Here the lattice depth is measured in units
of the recoil energy (parameter $s$). Due to the exponential
dependence on the square of the band gap, we can completely
suppress Zener tunnelling ($s>3$) or precisely tune it in order to
realize a continuous output coupler ($0<s<2$). If the momentum
spread of the atoms is smaller than the width of the first
Brillouin zone a constant fraction of the atoms is released from
the lattice and falls along gravity, every time the ensemble
reaches the Bragg momentum. The output is expected to be pulsed.
In the second method the lattice is switched off instantaneously
(non-adiabatically) thus projecting the quasi momentum
distribution on the momentum distribution. After a free expansion
of the atoms, the spatial distribution of the atoms reveals the
momentum distribution of the atoms inside the optical lattice. A
similar result is achieved in a slightly modified method, where
the optical lattice is switched off adiabatically. For these
purposes the lattice depth is lowered to zero in about 50 $\mu$s,
a time scale longer than the oscillation period of the atoms in
each lattice site. Thereby, the quasi momentum distribution is
exactly mapped (and not projected) to the momentum distribution.
From the absorption image we can extract the momentum distribution
of the atomic cloud. For interferometric applications one is
interested in the time evolution of the quasi momentum. To these
purposes we have to repeat the procedure for various trapping
times and perform a frequency analysis of the data.

The interpretation of the atomic distribution after release from
the trap as an interference pattern might not be obvious,
especially when only a single peak in the momentum distribution is
observed. The reason for this is that the wave nature of the atoms
is only apparent in the structure of the Bloch states whereas the
equations of motion (Eqns.\,\ref{eq1} and \ref{eq2}) are
classical. However, in an pure quantum-mechanical description
based on Wannier-Stark states the interpretation is
straightforward \cite{Roati2004a}: the atoms occupy the stationary
states of the Wannier-Stark ladder and after release, the
outcoupled atoms of all states interfere during the expansion.

\begin{figure}
\begin{center}
\includegraphics[width=12cm]{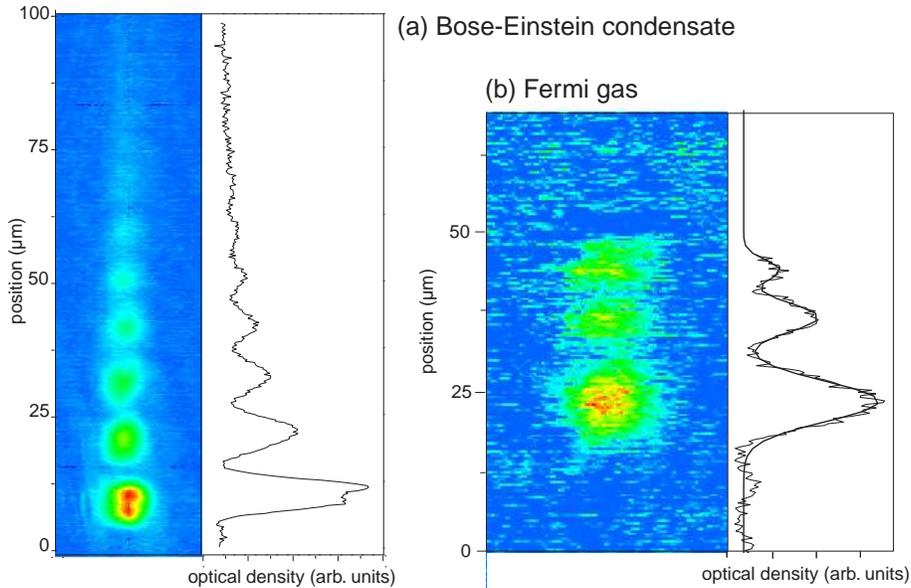} \caption{\label{fig3}
Absorption images after continuous release from the lattice. (a)
Bose-Einstein condensate: each pulse corresponds to a Bloch
oscillation. The first pulse has a total expansion time of 20 ms.
(b) Degenerate Fermi gas: due to the smaller atom number and the
shorter time of flight only three pulses are visible. The
expansion time of the first pulse was 8 ms. Both clouds have been
released immediately after being loaded in the vertical lattice
(zero evolution time). The scans on the right of both graphs
indicated the integrated optical density along the vertical
direction.}
\end{center}
\end{figure}

\section{Results and Discussion}

In Fig.\,\ref{fig3}a we show the interference pattern of a
Bose-Einstein condensate which is released continuously from the
lattice directly after being loaded. It shows the pulsed output
that is known for a coherent ensemble \cite{Anderson1998a}. The
distance between the pulses is defined by the Bloch oscillation
period which amounts to $T_{\mathrm{b}}=1.1$\,ms for the bosons
and $T_{\mathrm{f}}=2.3$\,ms for the fermions. The visibility of
the interference pattern for the condensate is large. Note, that
the vertical width of the pulses is much larger than the quasi
momentum spread of the condensate in the lattice. This is due to
the inter atomic repulsion that broadens the distribution during
the expansion. In Fig.\,\ref{fig3}b we repeat the experiment with
a degenerate Fermi gas. It shows the same structure however,
because of the longer Bloch oscillation period we observe only
three pulses. It is important to compare the visibility of the
pattern with that of the condensate: initially, the fermions have
a much broader quasi momentum distribution in the lattice. Its
half-width $\delta q$ is determined by the Fermi momentum
\cite{momentumspread} and for our parameters we find $\delta
q=0.75 q_{\mathrm{B}}$. This fulfills the requirement of a quasi
momentum distribution which is narrower than the recoil momentum
of the lattice. In contrast to the condensate, the momentum of the
fermions does not change during the expansion because the Pauli
principle forbids collisions between identical fermions at ultra
low temperatures. As a result the interference patterns for both,
the condensate and the Fermi gas show a similar contrast after the
expansion (Fig.\,\ref{fig3}).

\begin{figure}
\begin{center}
\includegraphics[width=8cm]{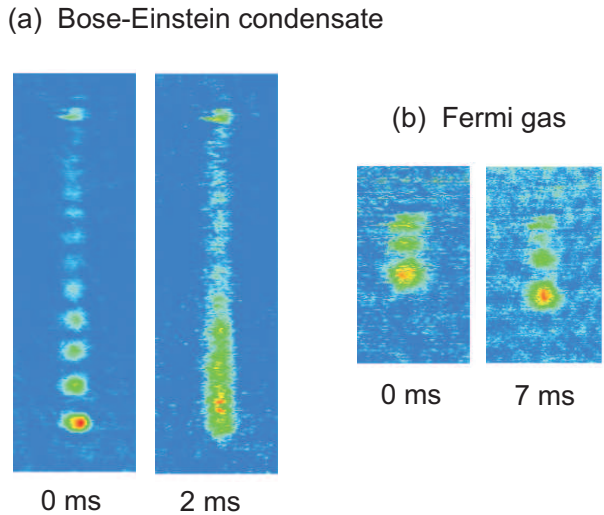} \caption{\label{fig4}
Absorption images after continuous release from the lattice for
different holding times. (a) Bose-Einstein condensate after 0 ms
and 2 ms of holding time, corresponding to about 1 Bloch
oscillations. (b) Degenerate Fermi gas after 0 ms and 7 ms,
corresponding to about 3 Bloch oscillations.}
\end{center}
\end{figure}

In a second experiment we have studied the evolution of the
interference pattern for longer holding times. After loading we
keep the atoms in the vertical lattice for several ms and repeat
the continuous release as described in the previous paragraph. The
results are shown in Fig.\,\ref{fig4}. While the distribution of
the fermions is unchanged with respect to the previous measurement
the interference pattern of the condensate has been washed out
completely. This striking difference is due to the interaction of
the atoms inside the condensate. A possible heating of the atoms
in the lattice can be excluded because the radial momentum
distribution does not change. We can give a qualitative
explanation of this effect by looking at the total energy
difference between atoms in neighboring lattice sites. In addition
to the gravitational potential the bosons experience a phase shift
due to the mean field energy. As the density profile of the
condensate along the lattice is inhomogeneous the contribution
from the mean field energy varies between neighboring lattice
sites. This leads to a dephasing of the wave function in different
lattice sites \cite{condensate} and to a loss of phase coherence.
In a spin polarized Fermi gas this effect is absent because
collisions between the atoms are forbidden. Hence, the phase
difference between atoms in neighboring lattice site is solely
determined by the external force. An important conclusion can be
drawn from this experiment: in interferometric applications of
trapped atomic gases it is more favorable to employ a degenerate
Fermi gas than a Bose condensed gas because the disadvantage of
having a relatively broad momentum distribution is overcompensated
by the absence of collisions.

\begin{figure}
\begin{center}
\includegraphics[width=8cm]{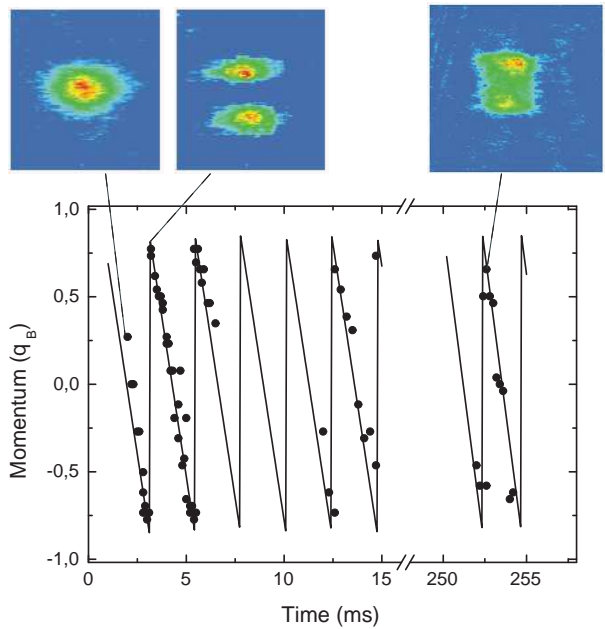} \caption{\label{fig5}
Time resolved Bloch oscillations. The graph shows the evolution of
the quasi momentum for 250 ms, corresponding to 110 periods. The
data are fitted with a sawtooth function. Absorption images are
shown for $t=0$, corresponding to the initial preparation of the
cloud at rest and at $t=T_{\mathrm{Bloch}}/2$, where half of the
cloud is Bragg reflected. The third image shows the cloud after
252 ms where the contrast has substantially decreased.}
\end{center}
\end{figure}

After having shown the superity of fermions for trapped atom
interferometry we now present an experiment where we have tested
the limits of an interferometry with trapped Fermi gases. Using
the same experimental sequence as described above we load a pure
Fermi gas in the optical lattice and follow the evolution of the
quasi momentum for many periods. At the end of each experimental
run we now adiabatically switch off the optical lattice and image
the cloud after 8 ms time of flight. From the absorption images
which show the quasi momentum distribution we can extract the
position of the main peak (Fig.\,\ref{fig5}). Indeed if we follow
its motion we find the peculiar sawtooth shape expected for Bloch
oscillations (see section \ref{section1}). We can follow the
oscillations for more than 250 ms corresponding to about 110 Bloch
periods, and only at later times the contrast is degraded by a
broadening of the momentum distribution. This is to our knowledge
the longest lived Bloch oscillator observed so far in all kind of
physical systems. The reduction of contrast can be seen from
Fig.\,\ref{fig5} where we show the absorption image of a data
point at 252 ms. The atoms fill nearly homogeneously the first
Brillouin zone and we cannot any longer determine the center of
the quasi momentum distribution with high precision. From a fit to
the present data we obtain a Bloch oscillation period of
$T_{\mathrm B}=2.32789(22)$ ms. Assuming that the only uniform
force acting on the atomic sample is gravity we determine a local
gravitational acceleration from Eqn.\,\ref{BO} as $g=9.7372(9)$
m/s$^2$.

The precision at the level of $10^{-4}$ obtained in this
proof-of-principle experiment is promising for future application
of this technique to accurate determination of forces. Note that
the extension of the Fermi gas in the vertical direction is very
small, $\Delta z=55\mu$m, and in principle can be reduced down to
the size of a few lattice sites without loosing the interference
pattern. Such a sensor based on trapped fermions is therefore
promising to extend the measurement of forces to length scales in
the micrometer range, with applications to the study of forces
close to surfaces, or to gravity at small scales
\cite{Dimopoulos}.

\section{Summary}
In conclusion we have studied the interference of trapped bosons
and fermions in an optical lattice aligned along gravity. For a
condensate interference can be seen with high contrast. However
the interference pattern is rapidly washed out for even short
evolution times in the vertical lattice. The absence of
interactions in a spin polarized fermionic sample prevents a
reduction of the contrast for longer evolution times and proves
the superiorness of fermions for trapped atom interferometry. We
have also demonstrated that interferometry with trapped fermions
allows for a high precision measurement of forces on a micrometer
length scale, with possible application to the study of
fundamental phenomena.

{\bf Acknowledgement} This work was supported by MIUR, by EU under
Contract No. HPRICT1999-00111, and by INFM, PRA "Photonmatter".
H.O. was supported by EU with a Marie Curie Individual Fellowship
under Contract No. HPMF-CT-2002-01958.


\begin{thebibliography}{77}
\bibitem{Berman1997a}
\textit{Atom Interferometry}, P. Berman (Ed.), Academic Press,
Chestnut Hill, MA (1997).
 (1997).
\bibitem{Peters1999a}
A. Peters, K. Y. Chung, and S. Chu, Nature \textbf{400}, 849
(1999).
\bibitem{McGuirk2002a}
J. M. McGuirk, G. T. Foster, J. B. Fixler, M. J. Snadden, and M.
A. Kasevich, Phys. Rev. A \textbf{65}, 033608 (2002).
\bibitem{Gupta2002a}
S. Gupta, K. Dieckmann, Z. Hadzibabic, and D. E. Pritchard, Phys.
Rev. Lett. \textbf{89}, 140401 (2002).
\bibitem{Anderson1998a}
B. P. Anderson and M. A. Kasevich, Science \textbf{282}, 1686
(1998).
\bibitem{BenDahan} M. Ben Dahan, E. Peik, J. Reichel, Y.
Castin, and C. Salomon, Phys. rev. Lett. \textbf{76}, 4508 (1996).
\bibitem{Ashcroft}
\textit{Solid State Physics}, N. W. Ashcroft and N. D. Mermin,
Saunders College, Phi (1976).

\bibitem{Roati2004a}
G. Roati, E. de Mirandes, F. Ferlaino, H. Ott, G. Modugno, and M.
Inguscio, Phys. Rev. Lett. \textbf{92}, 230402 (2004).
\bibitem{Roati2002a}
G. Roati, F. Riboli, G. Modugno, and M. Inguscio, Phys. Rev. Lett.
\textbf{89}, 150403 (2002).



\bibitem{stationary}
In an infinite system these states are not square-integrable and
are therefore often labeled as quasi stationary states or
resonances.
\bibitem{separation}
This is not strictly true because the lattice height varies across
the beam diameter, thus coupling the motion along the lattice with
the radial motion.
\bibitem{momentumspread}
Care has to be taken by calculating the momentum spread of a Fermi
gas in the direction of the lattice. Because of the
$\theta$-function shaped Fermi distribution the integration over
the two radial directions leads to a momentum spread which is
effectively smaller than the Fermi momentum.
\bibitem{condensate}
Because a tilted periodic potential does not posses an abolute
potential minimum one cannot define a ground state of the system.
Consequently a condensate which is loaded into such kind of
potential is never in equilibrium and cannot be described with a
single stationary wave function. In fact the experimental
observation directly proves this non-stationarity.
\bibitem{Dimopoulos}
S. Dimopoulos and A. A. Geraci, Phys. Rev. D \textbf{68}, 124021
(2003).

\end{thebibliography}
\end{document}